# Robust diffusion imaging framework for clinical studies

Ivan I. Maximov[1,*,*], Farida Grinberg[1], Irene Neuner[1,2,4], N. Jon Shah[1,3,4]

[1] Institute of Neuroscience and Medicine – 4, Forschungszentrum Juelich GmbH, 52425 Juelich, Germany

[2] Department of Psychiatry, Psychotherapy and Psychosomatics, RWTH Aachen University, 52074 Aachen, Germany

[3] Department of Neurology, RWTH Aachen University, 52074 Aachen, Germany

[4] JARA – BRAIN-Translational Medicine, RWTH Aachen University, 52074 Aachen, Germany

[*] Present address: Experimental Physics III, TU Dortmund University, 44221 Dortmund, Germany

[*] Corresponding author:

Ivan I. Maximov,

Institute of Neuroscience and Medicine – 4, Forschungszentrum Juelich GmbH,

52425 Juelich, Germany

Phone: +49 2461 61 8969

Fax: +49 2461 61 2820

Email: ivan@e3.physik.tu-dortmund.de

**Abstract**

Clinical diffusion imaging requires short acquisition times and good image quality to permit its use in various medical applications. In turn, these demands require the development of a robust and efficient post-processing framework in order to guarantee useful and reliable results. However, multiple artefacts abound in *in vivo* measurements; from either subject such as cardiac pulsation, bulk head motion, respiratory motion and involuntary tics and tremor, or imaging hardware related problems, such as table vibrations, etc. These artefacts can severely degrade the resulting images and render diffusion analysis difficult or impossible. In order to overcome these problems, we developed a robust and efficient framework enabling the use of initially corrupted images from a clinical study. At the heart of this framework is an improved least trimmed squares diffusion tensor estimation algorithm that works well with severely degraded datasets with low signal-to-noise ratio. This approach has been compared with other diffusion imaging post-processing algorithms using simulations and *in vivo* experiments. Exploiting track-based spatial statistics analysis, we demonstrate that corrupted datasets can be restored and reused in further clinical studies rather than being discarded due to poor quality. The developed robust framework is shown to exhibit a high efficiency and accuracy and can, in principle, be exploited in other MR studies where artefact/outlier suppression is needed.

**Introduction**

Investigating white matter organisation in the human brain with invasive surgical techniques is now rare because of the development of non-invasive medical imaging techniques (Budde and Frank, 2012). Diffusion tensor imaging (DTI) is a well established technique that can be used to detect the random displacement of water molecules on the micrometer scale by measuring the signal attenuation of diffusing molecules (Basser et al., 1994) and to reconstruct white matter architecture from this measured self-diffusion. More generally, diffusion imaging has been successfully applied in numerous neurological studies and diagnoses such as acute stroke, tumours and other disorders (Beaulieu, 2002; Johansen-Berg and Behrens, 2009; Jones, 2011; Le Bihan, 2007). Conventional DTI typically uses maps of scalar diffusion metrics, obtained from estimated diffusion tensor, such as mean diffusivity (MD), fractional anisotropy (FA), colour-coded FA, axial/radial diffusivity (L1/RD) and many other rotational invariants. In turn, the efficacy and reliability of diagnosis with DTI are strongly dependent on the workflow, experimental settings (shimming, pulse sequence, gradient encoding scheme, etc.) and post-processing methods (motion/eddy current corrections, noise correction, preselected estimation algorithms etc) (see recent review, Jones et al., 2012).

*In vivo*, the measurement of signal attenuation due to self-diffusion can be impaired by imaging artefacts caused by head motion, cardiac pulsation, respiratory motion and other effects. Hardware imperfections such as table vibrations caused by switching of magnetic gradient fields are an additional source of artefact. As a consequence, the measured dataset can be strongly corrupted and can exhibit large signal dropouts, also known as outliers. In turn, diffusion tensor evaluation becomes unstable and frequently leads to erroneous values. During the last decade, many different approaches have been introduced (Chang et al., 2005; Chang, 2010; Chang et al., 2012; Mangin et al., 2002; Maximov et al., 2011a; Maximov et al, 2011b; Niethammer et al., 2007; Pannek et al., 2012; Sharman et al., 2011; Zhou et al., 2011; Zwiers, 2010) in order to exclude artefacts from the diffusion signal attenuation and to improve the accuracy of tensor estimation. These proposed algorithms, such as RESTORE (Chang et al., 2005), least median squares (LMS) or least trimmed squares (LTS) (Maximov et al., 2011a) and PATCH (Zwiers, 2010), are well conditioned, either in the case of high signal-to-noise ratio (SNR) or when a redundant dataset is measured. Indeed, the original RESTORE algorithm may experience a problem excluding too many data points, especially, in the case of low SNR when "bad" data points can dominate over the "good". Under some circumstances, this may give rise to an unbalanced, or ill-defined, *b*-matrix (Chang et al., 2012). Another problem of the original RESTORE method is the absence of reasonable constraints in order to guarantee a positive definite diffusion tensor (Maximov et al., 2011a).

In turn, the LMS approach is very sensitive to the SNR and noise correction scheme, in particular, in the case of high diffusion weightings and low data redundancy (Maximov et al., 2011a; Maximov et al., 2012a; Maximov et al., 2012b; Maximov et al., 2015). The PATCH approach is potentially powerful and robust due to the use of a more sophisticated weighting function than the ordinary least squares algorithm (Zwiers, 2010). However, PATCH might also experience a problem with low SNR images, particularly in low redundancy datasets. It should be noted that in the case of low SNR, in particular, if a parallel imaging technique is being applied, the noise distribution becomes spatially variable (Koay et al., 2006; Landman et al., 2009a; Landman et al., 2009b; Manjon et al., 2010; Maximov et al., 2012a) and demands a proper noise correction approach (Maximov et al., 2012a; Rajan et al., 2012; Veraart et al., 2012).

Clinical DTI measurements are often restricted by limited scan times. Therefore, clinicians need simple and fast protocols for DTI studies. However, in extreme cases these protocols can yield substantially corrupted datasets that cannot be used for conventional DTI analysis. One such example is the measurement of Tourette patients (see Figure 1), where movement accompanying the tics of the patient cause data distortion.

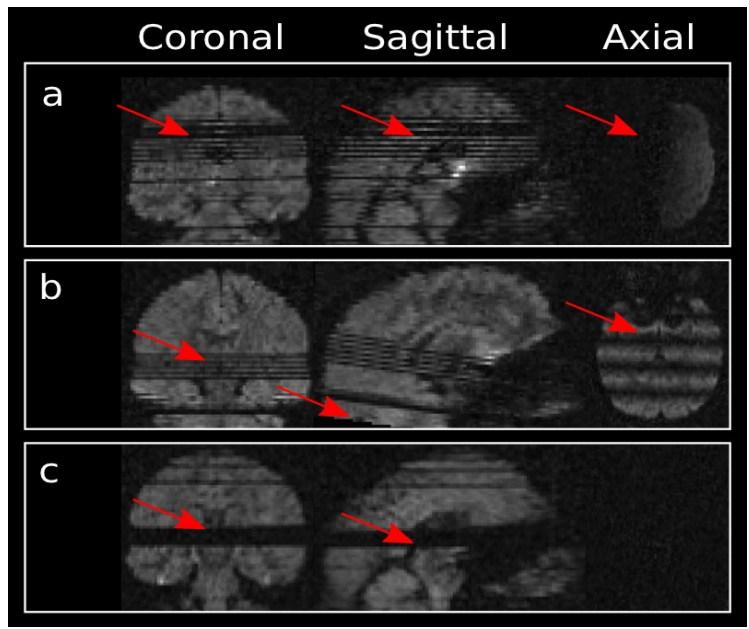

**Figure 1.** Examples of data distortions resulting from the tics exhibited by Tourette patients. a), b), c) show data from 3 different Tourette patients that show problems with data distortion (particularly severe are the errors in the axial slices and in patient c). Red arrows indicate the problematic regions.

Due to different movements that originated from Tourette-based motor tics, some slices of the whole-brain images can be corrupted or even completely distorted (see, in particular, Fig. 1a,c). Such types of distortions originating from the various artefacts mentioned above were also demonstrated in other studies (Gallichan et al., 2010; Niethammer et al., 2007; Pannek et al., 2012; Sharman et al., 2011; Walker et al., 2011; Zhou et al., 2011; Zwiers, 2010).

In clinical studies, data with severe distortions are excluded from analysis, and new measurements should be performed. However, it may happen that due to some unfavourable circumstances (patient's health condition, patient with a rare disorder or pathology, long term queues in MR centres, etc.) new measurements cannot be easily repeated. Recently, Chang et al., 2012, introduced a method based on the RESTORE algorithm, called "*informed* RESTORE", which allows one to analyse small data samples with significant signal dropouts. In the chemometric literature, the problem of small data samples corrupted by outliers is well known (Pison et al., 2002; Rocke, 1986; Rousseeuw and Verboven, 2002). Unfortunately, methods that analyse defective small samples were developed using only linear regression that are often not very useful in diffusion imaging (Jones et al., 2012). Initial studies using the LTS estimator have shown potential in application to small data sets (Maximov et al., 2012b). There are few publications dealing with the non-linear LTS regression generally (Cizek, 2002; Cizek, 2006; Cizek, 2008; Hawkins and Khan, 2009). Therefore we would like to introduce the novel and powerful non-linear robust estimator based on LTS approach to the MR community.

In this manuscript we propose a practical framework for post-processing highly corrupted diffusion datasets where standard methods fail to provide clinically useful results. A novel approach is presented that is based on the previously developed non-linear LTS estimator (Maximov et al., 2011a), which removes artefacts, of various origins, in low redundancy diffusion datasets with low SNR where diffusion tensor contrast is expected to be dependent on the noise level (Landman et al., 2008). The developed approach was compared with the well known non-linear least squares (NLS), constraint NLS using the Cholesky decomposition (CNLS), improved RESTORE, PATCH methods and the previous version of the LTS algorithm using simulations and *in vivo* experiments. The practical importance of the developed approach is demonstrated by the track-based spatial statistics analysis (Smith at al., 2006, 2007) based on a patient group with Tourette syndrome.

**Materials and Methods**

**Data acquisition**

Diffusion-weighted and high-resolution 3D $T_1$-weighted images were acquired for each subject on 1.5T Sonata Vision MR scanner (Siemens Medical Systems, Erlangen, Germany) with an 8-channel phased array head RF coil and a maximum gradient strength of 40 mT/m. The diffusion-weighted data were acquired using a twice-refocused spin-echo diffusion-weighted echo-planar imaging sequence with the following parameters: 2 mm slice thickness, no inter-slice gap, repetition time TR = 11000 ms, echo time TE = 89 ms, field-of-view FOV = 256 × 208 mm$^2$, imaging matrix = 128 × 104, number of slices in the transverse orientation = 71.

For each subject six images without diffusion weighting ($b = 0$ s/mm$^2$) and 30 images with non-collinear diffusion encoding gradients ($b = 800$ s/mm$^2$) were acquired. The time required for a single acquisition was 407 s. The acquisitions were repeated three times in order to improve the signal to noise ratio. $T_1$-weighted data were acquired using the standard Magnetization-Prepared, Rapid Acquisition Gradient-Echo (MPRAGE) sequence: TR = 2200 ms, inversion time TI = 1200 ms, TE = 3.93 ms, FOV = 256 × 256 mm$^2$, imaging

matrix = 256 × 256, number of slices = 128, flip angle = 15°, slice thickness = 1 mm, acquisition time = 578 s.

The Tourette patient and control group selection criteria were described in detail in Ref. (Neuner et al., 2010). All subjects gave prior written informed consent. *In vivo* measurements were divided into three groups: control group (CG) (22 patients), Tourette group without any significant visually-detected data distortions (TG1) (11 patients), and Tourette group with easy visually-detected data distortions (TG2) (11 patients). An example of typical data distortions in TG2 is presented in Figure 1.

**Theory**

In the following, we briefly describe the well known non-linear least squares algorithms with and without constraints, improved RESTORE, PATCH and the original and modified LTS approaches, respectively.

*Non-linear least squares (NLS and CNLS)*

An attenuation of the measured signal, $S_i$, in the *i*-th diffusion gradient direction is described as follows (Basser et al., 1994; Jones, 2011):

$$S_i = S_0 \exp(-b\, \mathbf{g}_i^T \mathbf{D}\, \mathbf{g}_i), \tag{1}$$

where $S_0$ is the signal amplitude without diffusion-weighting, *b* is the diffusion-weighting factor (the so-called *b*-value), $\mathbf{g}_i$ is the diffusion-encoding unit vector, and **D** is the second order, positive semidefinite symmetric diffusion tensor. In the diffusion experiment, measurements along at least six non-coplanar directions are required in order to resolve the six independent elements of the diffusion tensor. The target function in the case of non-linear least squares is as follows (Koay et al., 2006):

$$f_{\text{NLS}} = \tfrac{1}{2} \sum_{i=1}^{N} [S_i - S_0 \exp(-\mathbf{B}\, d)]^2, \tag{2}$$

where *N* is the number of applied diffusion gradients, $d = [D_{xx}\ D_{xy}\ D_{xz}\ D_{yy}\ D_{yz}\ D_{zz}]$ is the diffusion vector consisting of six independent elements of the diffusion tensor **D**, and **B** is the *b*-matrix (Jones, 2011) where each row of *b*-matrix corresponds to the given diffusion gradient **g** direction and *b*-value as follows $b \cdot [g_x^2,\ 2g_x g_y,\ 2g_x g_z,\ g_y^2,\ 2g_y g_z,\ g_z^2]$.

At the same time, a diffusion tensor should be a positive definite matrix, i.e. all eigenvalues of the diffusion tensor are positive. Cholesky decomposition of **D** permits the substitution of the variables, *d*, in Eq. (2) (Jones, 2011; Koay et al., 2006) and guarantees this property for the least squares algorithm. Otherwise a time-consuming constrained minimisation problem would be required to ensure the positive definiteness of **D**. In this manuscript, we used the Cholesky representation of the diffusion vector *d*; the corresponding function has a notation $f_{\text{CNLS}}$, where the Cholesky decomposition is as follows:

$\mathbf{D} = \mathbf{U}^T\mathbf{U}$, where $\mathbf{U}$ is the upper triangle matrix, with positive diagonal elements.

We have to note that Cholesky decomposition guarantees the positive semidefinite tensor $\mathbf{D}$ only in the case that all diagonal elements of the matrix $\mathbf{U}$ are positive.

*Improved RESTORE*

The original RESTORE algorithm utilises the weighted least squares approach where weightings are evaluated by the robust *M*-estimator (Chang et al., 2005). RESTORE excludes suspicious outlier values from further consideration using a reduced $\chi^2$ criterion. However, by removing the suspicious outlier values, RESTORE introduces two potential problems: removal of too many data from the fitting procedure and to yield an unbalanced or ill-defined *b*-matrix. In order to overcome these problems, Chang et al., 2012, improved the algorithm by adding criteria for a well-defined *b*-matrix that are the sum of the vector projections from the suspicious direction to the reference gradient sampling scheme and a condition number for *b*-matrix (Skare et al., 2000). The sum of vector projections is constant and should not depend on the gradient direction. As a criterion to stop removing the outliers, Chang et al., 2012, proposed the use of a redundancy coefficient (RC) that provides a threshold at one third of the sum of the vector projections. In our implementation of the improved RESTORE algorithm we used RC = 6 for 30 gradient directions, RC = 12 for 60 gradient directions, and RC = 24 for 90 gradient directions (diffusion encoding gradient scheme is JONES30, Jones et al., 1999). The detailed workflow of the algorithm is presented in Ref. (Chang et al., 2012). In this manuscript we used the improved RESTORE approach due to its generality and independence on type of signal distortion. This is in contrast to the informed RESTORE method where authors assumed the presence of decreased signal dropouts only.

*PATCH algorithm*

In order to improve the spatial detection of the outliers, Zwiers, 2010, suggested applying a more sophisticated weighting function consisting of three terms that account for: 1) cardiac motion artefacts, 2) head motions artefacts and 3) the noise correction originating from the linearization of Eq. (2). Weighting factors are, respectively, $w_i = w_{1i}w_{2i}w_{3i}$, where:

$$w_{1i} = \exp(- [A_1\, r_i/C_1]^2); \qquad w_{2i} = \exp(- [A_2\, E_i/C_2]^2); \qquad w_{3i} = S_i/\sigma_i. \qquad (3)$$

In Eq. (3), $A_1 = 0.3$ is the custom constant, $r_i$ are the residuals obtained from the linearised Eq. (2), $C_1$ is a spatially smoothed version of the median absolute deviation (MAD) estimator ($C_{MAD}$ = 1.4826 × median($|\,r_i\,|$) in the case of normally distributed data), $A_2 = 0.1$ is the custom constant, $E_i$ are the weighted residuals over all in-plane voxels, $C_2$ is the MAD estimator for $E_i$ data, $S_i$ is the expected signal and $\sigma_i$ is the expected noise level, respectively. An application of composite weightings in the linear iteratively reweighted least squares algorithm improves the diffusion tensor estimation. In turn, by setting $w_{i2}$ to unity, this approach can be easily reduced to the linearised RESTORE-like approach for a single voxel where instead of the Gelman-McClure M-estimator one has to use the Welsch function (Zwiers, 2010).

We have to notice that weightings $w_{1i}$ and $w_{2i}$ are estimated in the raw space after co-registration using affine transformation. Any deviations from the original algorithmic steps can lead to incorrectly estimated weightings. For example, simulations based on a single voxel model (one cannot use proper weighting functions at all), datasets after averaging (averaging procedure hides and distributes outliers over the brain volume), estimation of weighting functions in the reference space instead of raw space etc. can lead to incorrectly estimated weightings. In order to use the original PATCH algorithm we refer readers to the SPM implementation of the PATCH toolbox (Zwiers, 2010), available freely by request to author.

*LTS algorithm*

In the case of linear regressions, the LTS algorithm has been proven to be an efficient and robust estimator, particularly when substituted in place of the least median squares (Rousseeuw, 1984; Rousseeuw and Leroy, 1987). Recent adoption of the LTS algorithm (Maximov et al., 2011a) to a non-linear regression in diffusion tensor estimation provides an effective and fast algorithm for the outlier detection and removal. The LTS algorithm is based on the restriction of summation of the ordered residuals in conventional NLS algorithm. It can be formulated as follows:

$$f_{\text{LTS}} = \Sigma^{h}_{i=1}(r_i)^2, \qquad (4)$$

where $r_i$ are ordered $r_1 < r_2 < ... < r_N$ residuals of Eq. (2), and $h$ is the truncation factor (Maximov et al., 2011a).

*Modified LTS algorithm (MLTS)*

In order to increase the reliability of the robust estimations in the low redundancy datasets with low SNR we reorder the residuals $r_i$ using their absolute deviations from the median residual, i.e.

$$v_i = (r_i - \text{median}(r_j))^2, \qquad (5)$$

where $r_{i,j}$ are the residuals from Eq. (2), and $v_i$ are new reordering parameters. The reordering of the residuals taking into account the deviation from median residual allows one substantially to decrease an influence of multiple outliers in datasets with low redundancy due to less sensitivity of truncation in NLS to residuals with high/low values. The next evaluation with the truncation factor $h$ is the same as in the original LTS algorithm, i.e.

$$f_{\text{MLTS}} = \Sigma^{h}_{i=1}(v_i), \qquad (6)$$

where $v_i$ is ordered as $v_1 < v_2 < ... < v_N$. The details of the original and modified LTS algorithms are described in Appendix.

*Software*

All data were post-processed using in-house Matlab scripts (Matlab, The MathWorks, Natick, MA, USA). The motion/eddy current corrections were performed by the *eddy_correct* utility from the FSL package (Smith et al., 2004; Woolrich et al., 2009). Brain masks were evaluated using the BET utility (Smith, 2002) from the FSL package. The statistical analysis was done by the TBSS tool (Smith et al., 2006) from the FSL package. File conversion between the NifTI format and Matlab was performed with help of the Jimmy Shen Matlab scripts (http://www.rotman-baycrest.on.ca/~jimmy/NifTI/). For single slice diffusion tensor simulations and one subject *in vivo* measurements we used the version of PATCH implemented as a toolbox for SPM8 (Friston et al., 2007) courtesy of Dr. M. Zwiers.

*TBSS pipeline*

For all datasets we used one fixed pipeline. Initially, images in datasets were corrected for eddy-current induced distortions and related subject motions by FSL utility *eddy_correct* from FSL package (Smith et al., 2004; Woolrich et al., 2009). Appropriate *b*-matrix rotations were performed as well (Leemans and Jones, 2009). In order to improve the signal-to-noise ratio of datasets we used the power image procedure (McGibney and Smith, 1993; Miller and Joseph, 1993) with a noise correction estimated by the background algorithm (see, for example, Henkelmann, 1985). Using Brain Extraction Tool (BET) (Smith, 2002) we evaluated brain masks for each image in the datasets in order to exclude from consideration the regions consisting of no brain signals. The FA maps were estimated by all algorithms in each voxel using the previously extracted brain masks. After that, all subject FA images were aligned into Montreal Neurological Institute (MNI) 152 space using non-linear coregistration utitlity from FSL (FNIRT) (Smith et al., 2004). As a coregistration template FMRIB58_FA image was used. Next, a mean FA image was built up and thinned to produce a mean FA skeleton (Smith et al., 2006; Smith et al., 2007). An FA threshold equal to 0.2 was applied to the FA skeleton in order to exclude from consideration voxels which can be treated as grey matter or cerebrospinal fluids. Every subject FA image was projected onto this skeleton. Finally, datasets were prepared for voxelwise statistical analysis over the FA skeleton. Voxelwise analysis was performed using permutation-based, voxelwise non-parametric testing (Nichols and Holmes, 2002), implemented as "randomise" function in FSL package. The number of permutations in all tests was equal to 5000. A statistical thresholds $t > 3$, $p < 0.05$ and $p < 0.005$, corrected for multiple comparisons with threshold-free cluster enhancement (TFCE) were used during analysis (Smith and Nichols, 2009).

**Results**

Results are presented for the developed MLTS algorithm in comparison with the non-linear least squares, constrained non-linear least squares, improved RESTORE, original LTS and PATCH algorithms. The comparison is performed using the simulations and experimental *in vivo* diffusion data obtained for the Tourette patient study. In the simulations we used a single voxel model and a single slice synthetic diffusion dataset in order to emphasise the statistical distribution of the estimations for all approaches, in particular, for different noise levels and outlier contamination rates. We should note that distortions are only caused by artificial outliers in contrast to added thermal noise producing signal variations with a known noise distribution.

**Simulations**

*Single voxel model*

We assessed all methods using synthetic diffusion data for one voxel with 30 diffusion encoding gradients. The original diffusion tensor for all simulations had following parameters: eigenvalues [0.7; 1.1; 3.0] × $10^{-3}$ mm$^2$ s$^{-1}$, MD = 1.6 × $10^{-3}$ mm$^2$ s$^{-1}$, FA = 0.65066. Original signal attenuation was distorted by Rician noise with standard deviation running range [0.025; 0.175] with step 0.025 (7 points with corresponding SNR equals to 40, 20, 13.3, 10, 8, 6.6, and 5.7 or in a percentage of maximal signal at *b*-value = 0: [2.5; 5; 7.5; 10; 12.5; 15; 17.5] %). A number of outliers (0, 2, 4, 6, 8, 10, 12, and 16) were randomly distributed over 30 gradient directions. The amplitude of the applied outliers was randomly chosen from the range [0, 1] with uniform distribution of the random value. Diffusion weighting, *b* = 1000 s mm$^{-2}$, was equal for all simulations.

The results of these simulations are presented in Figure 2. The estimations of the diffusion tensor obtained by all algorithms are compared statistically. The statistical test consisted of 1000 simulated variants of outlier/noise distribution for each number of outliers, i.e. 0, 2, 4, 8, 10, 12, 14 and 16. The comparison criteria were based on the angle *α* (the angular deviation of the estimated main eigenvector from the original one), the L1, FA and MD values, and reduced $\chi^2$. Note that the reduced $\chi^2$ is not a perfect goodness-of-fit criterion, in particular, in the case of robust estimations. However, this parameter could provide additional information about the reliability of the estimations obtained by robust approaches comparing to conventional least squares and allows one to check reproducibility of the estimations by preliminary detection of leverage points in original datasets. We have to mention that evaluations at low SNR exhibit additional problems for the robust estimators due to significant signal variability and a weak criterion for outlier detection. All robust estimators experience a problem and can detect a true signal as an outlier. As a consequence, the resulting diffusion metrics can be biased (see Fig. 2 ).

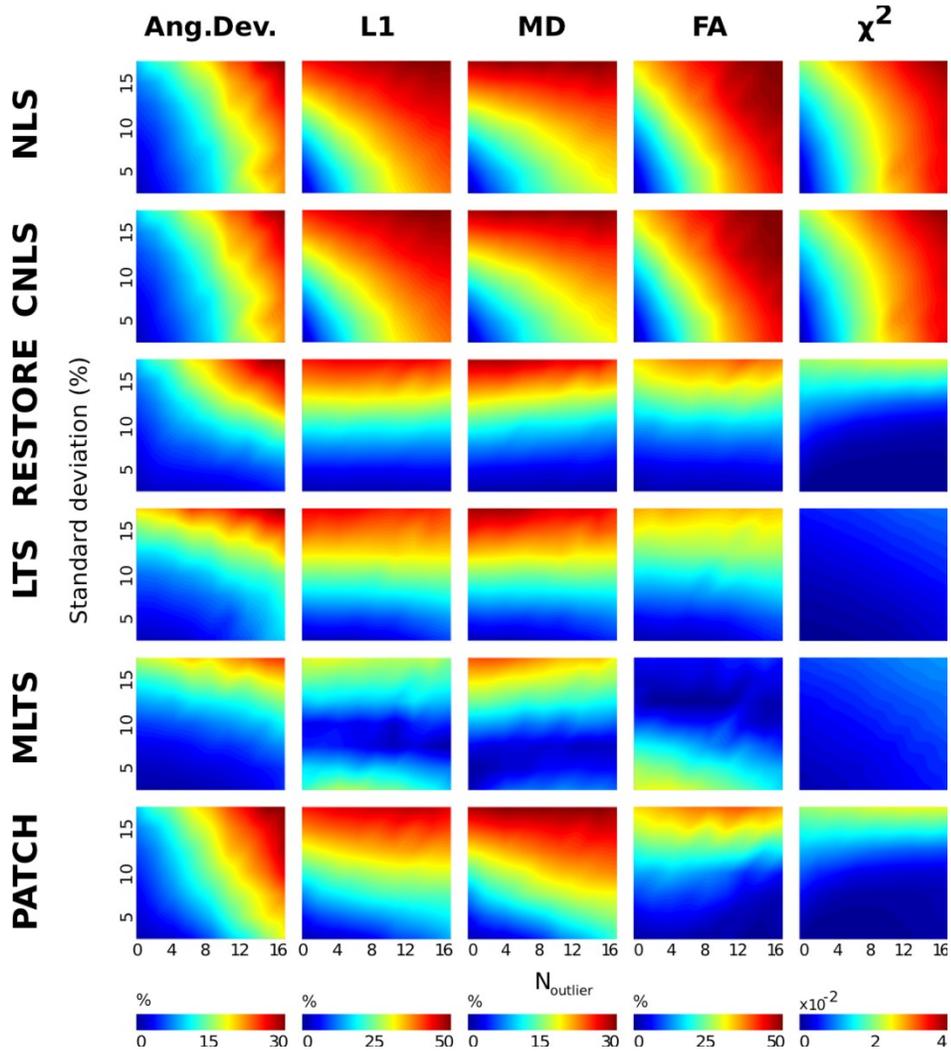

**Figure 2.** Results of a comparison between different algorithms of statistically simulated diffusion tensor estimations. The comparison was performed for standard deviation from the range [2.5; 5; 7.5; 10; 12.5; 15; 17.5]%. The original tensor values were L1= 3.0 × 10$^{-3}$ mm$^2$ s$^{-1}$, MD = 1.6 × 10$^{-3}$ mm$^2$ s$^{-1}$, FA = 0.65066 with an angular deviation $\alpha$ = 0°. The colour code corresponds to the mean deviation of estimated values comparing to the original values in percentage. The values for reduced $\chi^2$ are presented as is.

The results of comparison are presented in relative form as a deviation taken in percents from original values and averaged over statistical trials (1000 samples).

*Single slice synthetic diffusion dataset*

In order to demonstrate the efficacy of the PATCH algorithm based on a factorised weighting function, we performed a simulation of a single slice dataset consisting of 50×50 in-plane voxels. The number of diffusion gradients was selected equal to 30. The simulated diffusion dataset has five different regions: a background region with zero signal and Rayleigh distributed noise, a region with isotropic diffusion tensors "head" (FA = 0, MD = $1\times10^{-3}$ mm$^2$ s$^{-1}$), regions with anisotropic diffusion tensors: "left eye" (FA = 0.65, MD = $1.1\times10^{-3}$ mm$^2$ s$^{-1}$), "right eye" (FA = 0.68, MD = $1.4\times10^{-3}$ mm$^2$ s$^{-1}$), and a "mouth" region (FA = 0.21, MD = $0.8\times10^{-3}$ mm$^2$ s$^{-1}$). The regions with non-zero signal have been corrupted by the Rician noise with the resulting SNR of approximately 20. We randomly selected 6 diffusion gradients (5% contamination rate) and introduced outliers for these directions as follows: 3 diffusion directions have been distorted completely (see Fig. 3, Outlier 2), and 3 directions have been distorted by outliers only inside a preselected area (see Fig.3, Outlier 1 and yellow square at the Original image marked preselected region). The amplitude of the outliers was uniformly distributed over the range [0, 1].

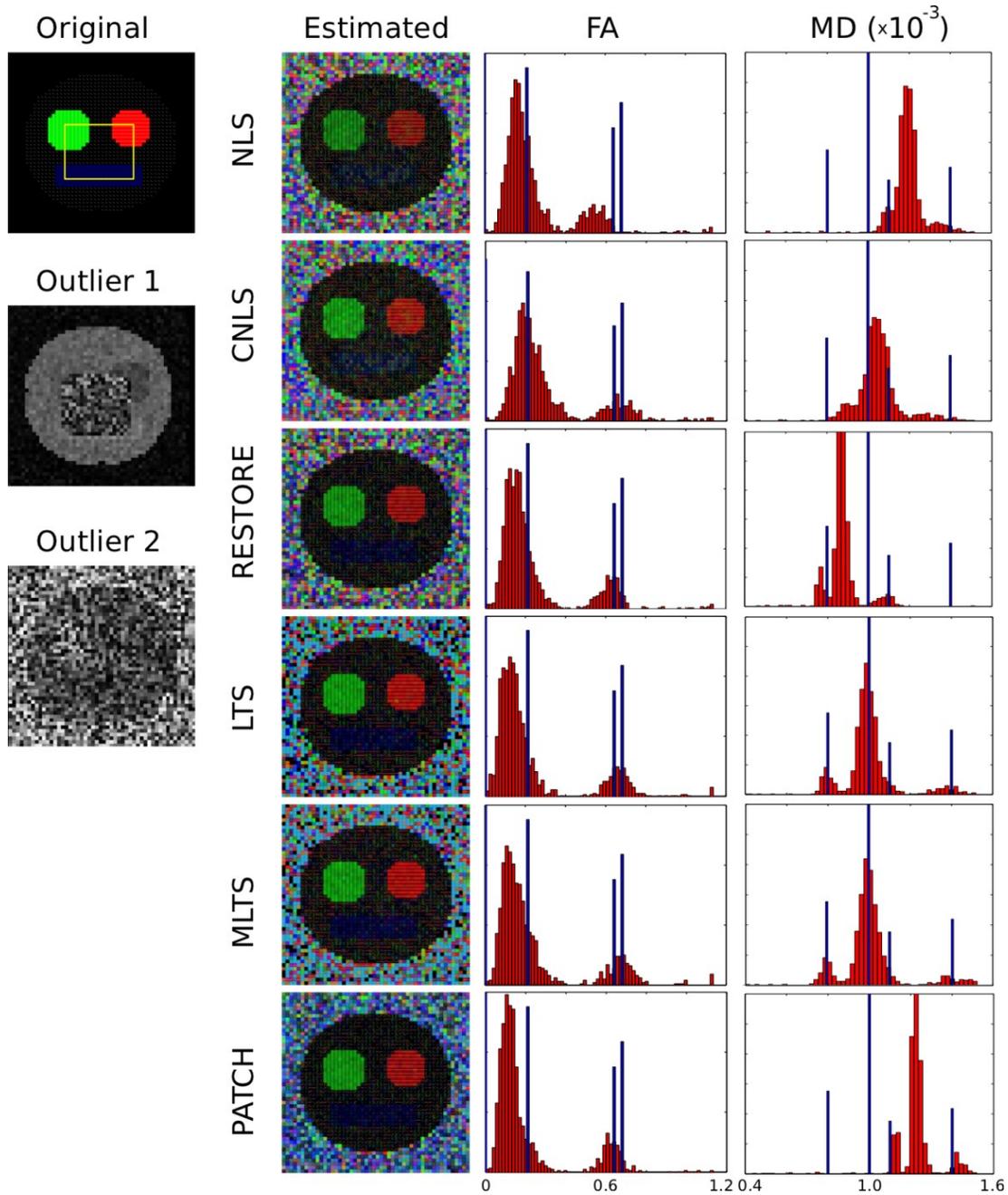

**Figure 3.** Evaluations of diffusion scalar metrics for the single slice synthetic dataset with 30 encoding gradients, obtained by all algorithms. A contamination rate was 5% (6 randomly selected diffusion directions). Two types of outliers were introduced: outlier 1 is inside a preselected region marked by yellow square frame (3 diffusion directions) and outlier 2 is complete image distortion. Resulting colour coded FA and histograms of FA and MD over the non-zero signal regions were computed. Blue bars in histograms present original values of FA and MD for the "head" (black region), "left eye" (green), "right eye" (red), and "mouth" (blue) regions.

In Fig. 3 we present colour coded FA estimated by all algorithms and histograms of FA and MD values estimated over non-zero signal regions. The original values of FA and MD are highlighted by blue bars with appropriate bar length.

*In vivo* **study: one subject**

In order to demonstrate the differences in diffusion tensor estimation of whole-brain measurements we compared all algorithms using a single subject. The original datasets were strongly distorted in each of the three acquisitions used to increase the SNR and decrease the statistical bias of estimations (see Figure 4: Original).

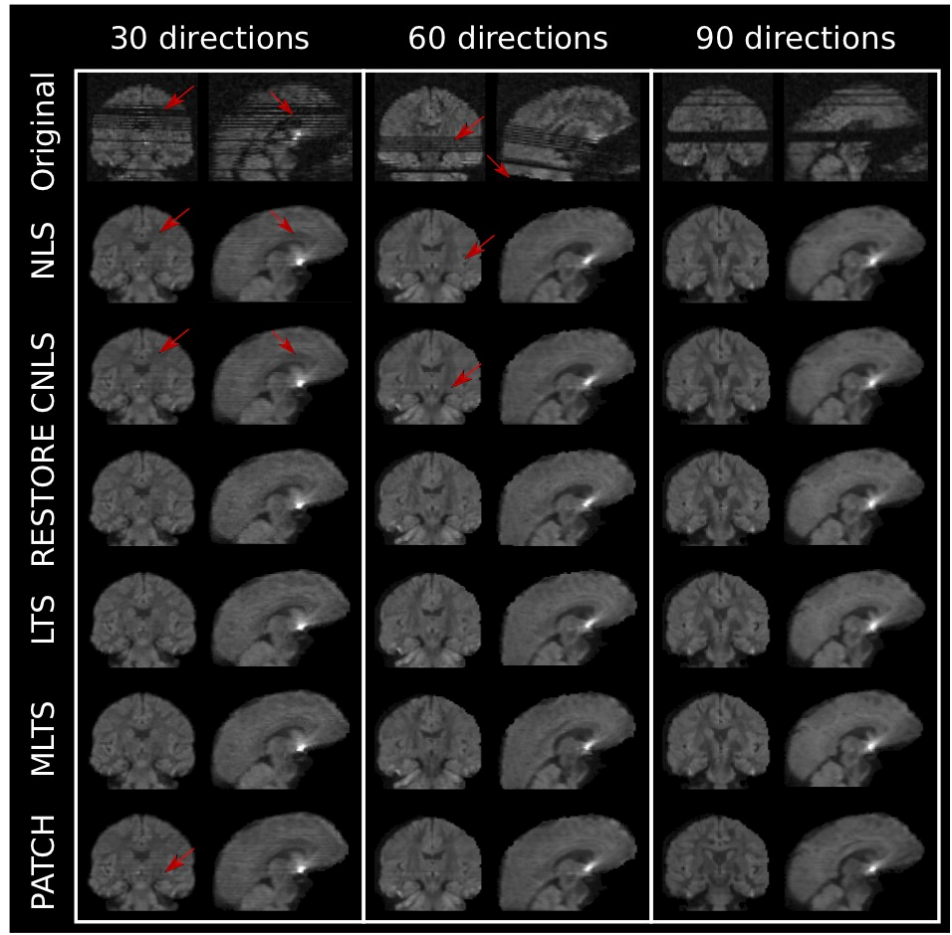

**Figure 4.** Three sets of DWI data obtained for the same subject with different numbers of encoding gradients. The first row is original DWI for preselected directions. Below the reconstructed DWI are presented with respect to the applied algorithms. The red arrows emphasize the regions where the residual artefacts can be detected due to degraded slices in original DWI.

In order to improve the estimation eddy currents/motion correction has been done using FSL for all approaches excepting PATCH. However, this procedure is strongly demanded due to multiple head motions originated by dramatic patient tics and tremor. As a result, in the case of 60 directions additional head rotation was a reason of extra artefacts (see, for example Fig. 4: Original image for 60 directions sagittal view).

The results of the statistical comparison between the six algorithms are represented by the scatter plots in Figure 5. In Fig. 5 the scatter plots were computed using the mean values of diffusion scalar metrics estimated over each slice. Every dataset was marked by different colour in order to emphasize the difference in evaluation with increasing number of encoding diffusion gradients: red points are 30 directions, blue points are 60 directions, and green points are 90 directions.

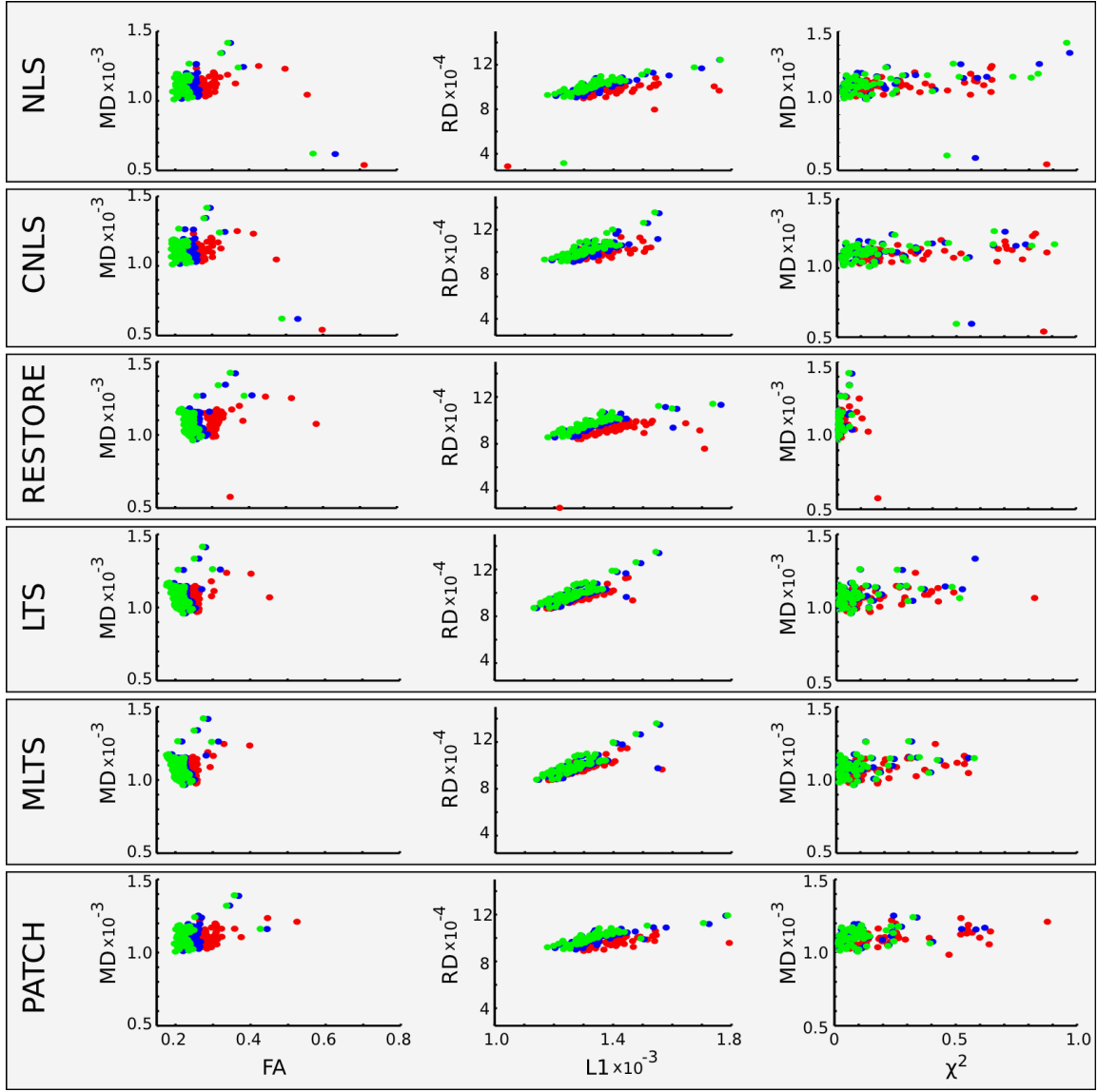

**Figure 5.** Comparison of results for all 6 algorithms using scatter plots. Each point on the scatter plot corresponds to mean value estimated over slice. Red colour corresponds to 30 direction dataset, blue corresponds to 60 direction dataset, and green corresponds to 90 direction dataset. The left column is scatter plot between FA and MD (mm$^2$ s$^{-1}$), the middle column corresponds to L1(mm$^2$ s$^{-1}$) versus RD (mm$^2$ s$^{-1}$), and the right column corresponds to $\chi^2$ versus MD(mm$^2$ s$^{-1}$).

The parameters of scatter plots in Fig. 5 can be presented in quantitative form using non-parametric correlation coefficients based on Spearman's $\rho$ parameter. The correlation coefficients for all scatter plots are summarized in Table 1:

**Table 1:** Spearman's correlation coefficients depending on the number of gradient directions for Fig. 4. P-values for all correlation coefficients are less than $10^{-5}$.

| Algorithm | | 30 vs 60 | 30 vs 90 | 60 vs 90 |
|---|---|---|---|---|
| NLS | FA | 0.63 | 0.58 | 0.86 |
|  | MD | 0.88 | 0.89 | 0.89 |
|  | L1 | 0.86 | 0.91 | 0.92 |
|  | RD | 0.83 | 0.83 | 0.83 |
| CNLS | FA | 0.64 | 0.57 | 0.85 |
|  | MD | 0.88 | 0.89 | 0.89 |
|  | L1 | 0.85 | 0.84 | 0.97 |
|  | RD | 0.89 | 0.89 | 0.89 |
| RESTORE | FA | 0.66 | 0.57 | 0.93 |
|  | MD | 0.94 | 0.92 | 0.92 |
|  | L1 | 0.91 | 0.89 | 0.97 |
|  | RD | 0.87 | 0.87 | 0.87 |
| LTS | FA | 0.78 | 0.74 | 0.97 |
|  | MD | 0.96 | 0.95 | 0.95 |
|  | L1 | 0.96 | 0.95 | 0.97 |
|  | RD | 0.96 | 0.96 | 0.96 |
| MLTS | FA | 0.79 | 0.77 | 0.98 |
|  | MD | 0.96 | 0.94 | 0.98 |
|  | L1 | 0.95 | 0.94 | 0.98 |
|  | RD | 0.96 | 0.97 | 0.97 |
| PATCH | FA | 0.64 | 0.59 | 0.87 |
|  | MD | 0.88 | 0.90 | 0.91 |
|  | L1 | 0.78 | 0.79 | 0.97 |
|  | RD | 0.81 | 0.81 | 0.82 |

Some of the points in Fig. 5 can be treated as "super"-outliers due to substantial deviation from the centre of scatter plots. One of the possible reasons of that is the motion correction which introduce additional artefacts, particularly, over the first and last slices. This effect is strongly pronounced in the case of NLS and CNLS algorithms (see Fig. 5). The noise level dependence of the estimated diffusion scalar metrics is emphasized in Table 1 where correlation coefficients are presented for pairs of directions such as 30 vs 60; 30 vs 90; and 60 vs 90. One can see that highest correlation values are exhibited for 60 vs 90 case.

In order to emphasize the differences in parameter estimations depending on number of directions we plotted Figure 6 with colour coded FA.

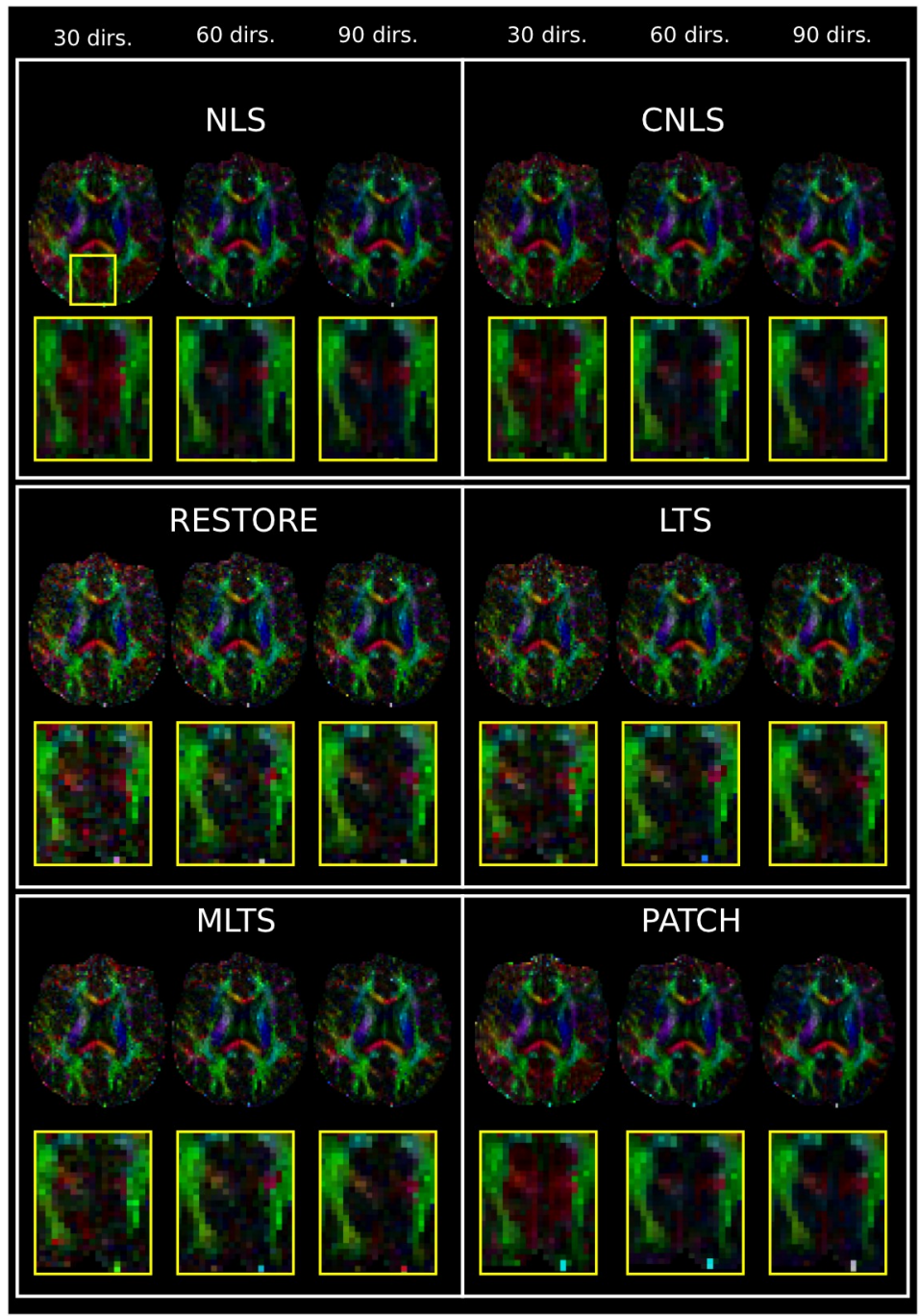

**Figure 6.** Colour-coded FA. In order to emphasize the difference in colour-coded FA estimation we enlarged the regions marked by yellow frame (see NLS subframe) for each algorithm.

Due to low SNR in the case of 30 directions the NLS, CNLS and PATCH could not properly estimate the diffusion scalar metrics and associated diffusion tensors (see Fig. 6). The images obtained by RESTORE, LTS and MLTS substantially decrease the artificial estimations but produce very noisy maps. The increased number of diffusion weighted directions allows

PATCH produce acceptable colour FA image already for 60 diffusion directions when NLS and CNLS still have artefacts in estimated diffusion metrics.

**_In vivo_ study: track-based spatial statistics analysis**

Conventional clinical DTI studies frequently use FA maps in order to localize various white matter changes, for example, related to degeneration processes or developmental disorders. One of the powerful approaches which allow one to detect such changes is represented by the TBSS tool from the FSL package (Smith et al., 2006; Smith et al., 2007). In the following, we use TBSS to localise the spatial changes of white matter metrics for the control (CG) and Tourette patient (TG2) groups.

First, we estimated FA maps applying the NLS algorithm to CG and TG1 groups. DWI images in the CG and TG1 groups have the best SNR (three acquisitions were applied). After that, we performed the TBSS analysis between CG and TG1 following a standard protocol (Smith et al., 2007) and partially (due to smaller number of participants in TG1 group; 11 patients in TG1 versus 19 patients in Ref. (Neuner et al., 2010)) reproduced the results from Ref. (Neuner et al., 2010). The results of this analysis are presented in Figure 7a where the Montreal Neurological Institute (MNI) $T_1$-weighted 152 brain image is used as a background. In Figure 6a, we demonstrated the regions with significantly ($p < 0.01$) decreased FA in Tourette patients with respect to controls. In order to emphasize these regions, we applied _tbss_fill_ utility from the TBSS tool. In this section, the image in Fig. 7a will be treated as a reference.

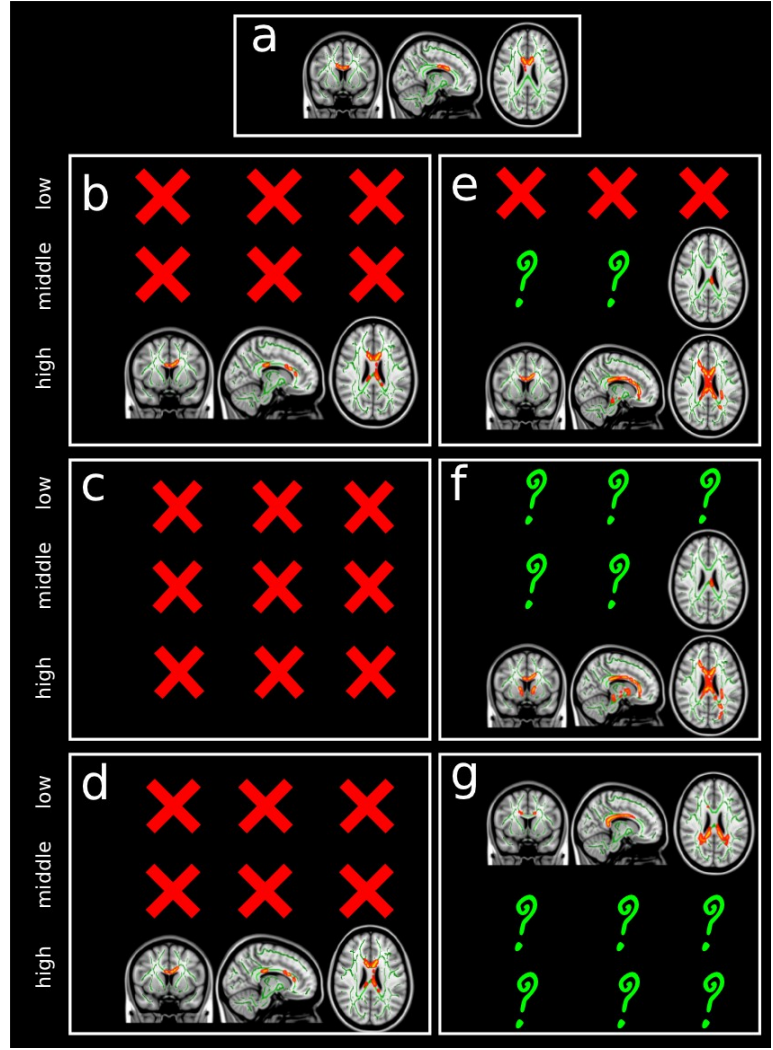

**Figure 7.** Results of TBSS FA analysis. a) TBSS for CG vs. TG1. b)-g) TBSS FA for CG vs. TG2 for different number of averages obtained using b) NLS, c) CNLS, d) RESTORE, e) LTS, f) MLTS and g) PATCH. The red cross means that no significant ($p < 0.01$) differences are found. The green question mark denotes the fact that no significant differences were found for the fixed MNI coordinates. MNI coordinates for all images are X = 11, Y = 8, Z = 24. The green colour is used for mean skeleton.

The TG2 group was evaluated taking into account three SNR as low, middle, and high. After that, we performed the same TBSS analysis as in the case of CG vs. TG1 with all SNRs in TG2 group. The results are presented in Fig. 7. In Fig. 7b, the NLS algorithm was used. Only in the case of high SNR could regions of significantly ($p < 0.01$) decreased FA in Tourette patients be detected with distorted images. In the case of low and middle SNR, the statistical analysis did not detect any regions with significant difference. In Fig. 7c, the CNLS algorithm was used for FA map estimations but no regions with decreased FA were found regardless of SNR. Results obtained using the RESTORE and LTS algorithms are shown in Figs. 7d and 6e respectively. In the case of low SNR, no regions with significant differences were found. For middle SNR, regions with decreased FA were found in Tourette patients using the LTS algorithm. However, these regions differ from those in the control analysis shown in Fig. 7a. For the data with high SNR, the same regions as found in the control data were detected. The MLTS algorithm yield results very similar to those produced with the LTS except when low SNR was used. With low SNR, regions with significant differences were identified but they do not coincide with the control analysis. The PATCH algorithm, in contrast to all others, yielded regions with significantly decreased FA in Tourette patients in the case of low SNR. However, the regions detected in the cases of middle and high SNR differ from those found in the control analysis.

Preliminary comparison of the results obtained by TBSS with the previous investigations (Neuner et al., 2010) indicate that, with high SNR data, LTS and MLTS exhibit similar detection of the regions with decreased FA in Tourette patients. In order to demonstrate advantages of the developed robust framework, we mixed up two groups: TG1 where FA maps were obtained using the NLS algorithm and TG2 where FA maps were obtained using the MLTS algorithm. We performed an additional TBSS analysis of the mixed TG group and CG. The results of this TBSS analysis are presented in Fig. 8. The criterion for a significant difference in this experiment was set more rigorously as $p < 0.005$. In both cases, similar regions were detected with decreased FA in the anterior and posterior limbs of the corpus callosum of the Tourette patients. However, unlike the results obtained in Ref. (Neuner et al., 2010), a significant difference was not found in the internal capsule. Instead, differences in the thalamus region of the TG group in comparison to the healthy controls can be observed.

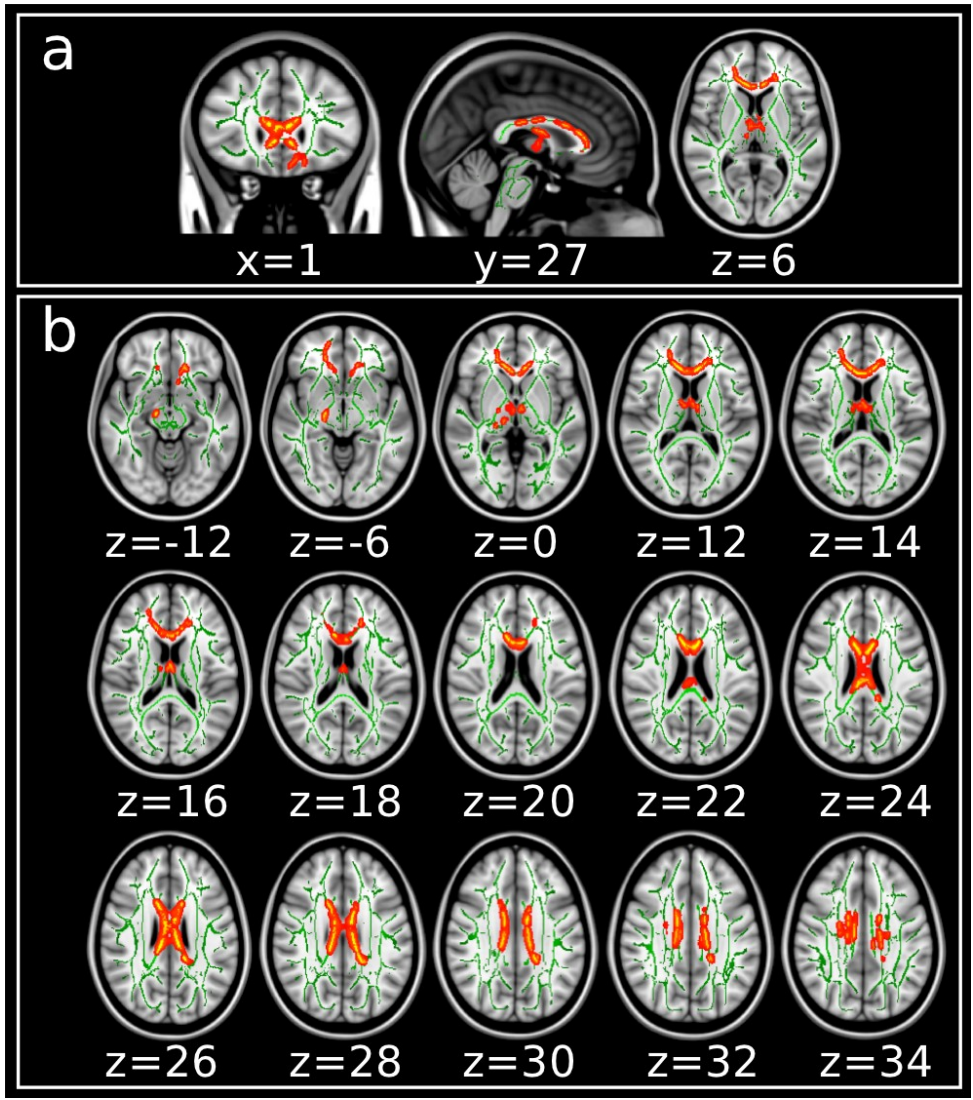

**Figure 8.** Results of TBSS FA analysis for the mixed group TG1+TG2/CG. The MNI coordinates in a) and b) are the same as in Ref. (Neuner et al., 2010; Fig. 1 and Fig. 2) and are given below of each image. The regions detected with significant difference ($p < 0.005$) using the *tbss_fill* utility are shown in red. The green colour is used for mean skeleton.

**Discussion**

In this paper, we addressed potential problems that frequently arise in clinical DTI studies due to a presence of artefacts. We described a robust post-processing framework, which permits processing of datasets contaminated by severe artefacts such as those caused by Tourette syndrome movement tics. Corrupted image data can be restored with this framework and included in studies where they would otherwise be discarded. Our approach exhibits improved results when compared to other methods as demonstrated by the TBSS analysis of *in vivo* measurements.

Conventional DTI tools that use robust estimators are well known and frequently used, in particular, RESTORE which is a recommended method (Jones et al., 2012). However, these methods can experience the problems when applied to problematic data such as data with low SNR, a small data samples, low redundancy of DTI measurements, severe artefacts caused by subject and/or table motion, etc. It can clearly be seen in Figs. 2 and 3 that, even in the case of simple simulations, these algorithms experience problems in accurately estimating diffusion tensors in the presence of noise and data errors. Note, that the uniform distribution of amplitudes of the outlier in statistical simulations allows us to cover the multiple effects of degraded signals, for example, such as residual artefacts after eddy current/susceptibility corrections. For high SNR, good results can be expected from the LTS, MLTS, and improved RESTORE algorithms, even in the cases where outliers corrupted almost half of the encoding gradient directions. What gives a nice coincidence with a breakdown criterion of selected robust estimators (Rosseeuw and Leroy, 1987). PATCH based only on the Welsch weighting function exhibits reasonable estimations for high SNR and not high number (approximately 20%) of outliers only in single voxel model. Increasing number of outliers and low SNR produce remarkable deviations in PATCH evaluation. Single slice simulations exhibit basic problems with conventional least squares approaches for dataset corrupted by different types of outliers. At the same time all robust algorithms produced qualitatively adequate images of colour-coded FA even in the case of anisotropic diffusion tensors. However, a quantitative analysis of the estimated FA and MD values demonstrated that only LTS and MLTS approaches could reconstruct a distribution which is close to true values. We have to note that in the single slice simulations we do not need motion/eddy current corrections.

Landman et al. compared various estimation methods and diffusion-weighting schemes with *in vivo* measurements of identical data with different signal-to-noise values and demonstrated the effect of the noise on tensor contrasts. (Landman et al., 2008). The related experiment that we performed with the same subject is more complicated due to additional artefacts originating from severe Tourette tics in which one or more slices of the data is completely corrupted. Even more problematic are measurements with a small number of applied diffusion gradients (as six or twelve, for example) where the problem becomes practically unsolvable. Here, we should emphasize once again that a choice of the diffusion-weighted scheme in clinical experiments can significantly influence the accuracy of, and introduce variability in, DTI rotational invariants in post-processing analysis (Tournier et al., 2011). Thus, the algorithms that can suppress artefacts (physiological noise, tics motions, noise related bias etc.) and increase the evaluation accuracy are desirable. By comparing the performance of several techniques applied to datasets with various SNRs, we would opt to use the robust estimators, particularly, the MLTS algorithm for several reasons. First of all, the robust-based algorithms provide a stable estimation of the diffusion tensor under different conditions as demonstrated in Figs. 5,6 and Table 1. Note that an increasing number of diffusion gradients (see Fig. 5 and Table 1) allows one to see a statistical changes of diffusion metrics distributions in the scatter plots and correlation coefficients. In the case of robust estimators such as RESTORE,

LTS/MLTS, and PATCH we expect to have similar value distributions for the maximal number of diffusion gradients in contrast to the conventional least squares approaches due to independence of the robust estimations on the presence of multiple artefacts. At the same time, quantitative comparison of the obtained results, such as colour-coded FA images (see Fig. 6), allows us to provide a conventional presentation of diffusion metrics estimated by all algorithms. In this study, the algorithms based on robust estimators, in particular, the MLTS could be treated as reliable and recommended for a practical use with dataset contaminated by multiple artefacts. We should note that the recently developed *informed* RESTORE algorithm (Chang et al., 2012) is potentially comparable in terms of accuracy and efficiency to presented MLTS. We plan to perform rigorous comparisons of *informed* RESTORE with MLTS in future work. However, we have to emphasize that *informed* RESTORE exhibits all of these benefits in the case of dramatically decreasing signal only. In general cases, improved RESTORE seems to be a more appropriate approach, particularly, for the voxel-based analysis.

Interesting results related to the statistical analysis of Tourette and healthy control groups were obtained. Recent publications (Neuner et al., 2010; Neuner et al., 2011) based on artefact-free data revealed that Tourette patients exhibit decreased FA values in parts of the corpus callosum (anterior and posterior body), which were identified by the TBSS tool. On the other hand, we know that many voxelwise analyses are criticised because of the effect of artefacts and non-linear coregistration procedures on the data (Smith et al., 2007). Moreover, different voxelwise approaches reproduce significantly different results over the same target group (Jones et al., 2007). In order to perform a fair TBSS comparison of all algorithms we used one statistical tool for the same groups. It means that, since all parameters are identical during TBSS workflow, differences in the TBSS results can be attributed to the algorithm selection only. The results of TBSS analysis exhibit improved reliability of the MLTS with respect to the other algorithms for datasets strongly contaminated by artefacts such as those created by Tourette-based motion tics. The regions of substantial difference in FA estimated by the MTLS algorithm were more close to the control group test regions than those obtained by the other algorithms (see Fig 7). In order to corroborate this effect we combined the MLTS corrected TG2 group data with the TG1 data and performed the same analysis as presented in Ref. (Neuner et al., 2010). The results of the mixed-group TBSS analysis showed a satisfactory agreement with those presented in Ref. (Neuner et al., 2010). By including the initially corrupted (but restored) images into the study, we succeeded in doubling the target group of Tourette patients taken into consideration and thereby demonstrate the efficiency and potential applicability of the proposed framework to future clinical studies.

**Conclusions**

The developed robust post-processing framework is based on a modified least trimmed squares method and exhibits good performance in clinical datasets. Most notably, it reliably provides results with improved precision and data efficacy. In practice, the proposed diffusion imaging framework permits the inclusion of data that has been severely degraded by various artefacts in diffusion studies, such as track-based spatial statistics, where previously data would have been necessarily discarded. It is hoped that the proposed approach might find other applications in MRI where datasets are contaminated by outliers. As an example, our approach might be useful for estimation of other diffusion parameters in non-Gaussian or high angular resolution models where typically interplay between SNR and abundance comes in front of the problem.

**Acknowledgements**

IIM thanks Dr. Marcel P. Zwiers who brought our attention to the Ref. (Zwiers, 2010) and kindly shared his PATCH code with us and Dr. Pavel Cizek who shared with us his publications. Authors thank Dr. Tony Stoecker for valuable discussions and Dr. Michael Poole for proof reading of the manuscript and useful suggestions improving the text.

**Appendix**

In this appendix we provide details of the MLTS algorithm based framework of which a flow diagram is presented in Fig 9. The first step in the algorithm was to visually check all acquisitions for any potential artefacts. Secondly, noise correction and averaging of all acquisitions was performed in order to improve SNR. This step typically involves eddy current and simple motion corrections as well. Note that motion correction should be done in a relation to proper *b*-matrix correction (Leemans and Jones, 2009). In the next step, a brain mask is prepared and a loop is performed over all brain voxels. Initial signal attenuation fitting is performed with the help of the non-linear least squares. The criterion for a poor fit or potential outlier existence can be selected in a different way (Chang et al., 2012; Maximov et al., 2011; Maximov etl al., 2015). We used the method described in Ref. (Chang et al., 2012) as a robust and reliable one. If the criterion is not met the algorithm directly proceeds to the next voxel. Otherwise, if the criterion is met, the MLTS algorithm was applied, which is described in Ref. (Maximov et al., 2011). However, instead of Eq. (4) we use Eq. (6). In order to exclude processing of the negatively defined diffusion tensors in the MLTS algorithm Cholesky decomposition was used (Koay et al., 2006).

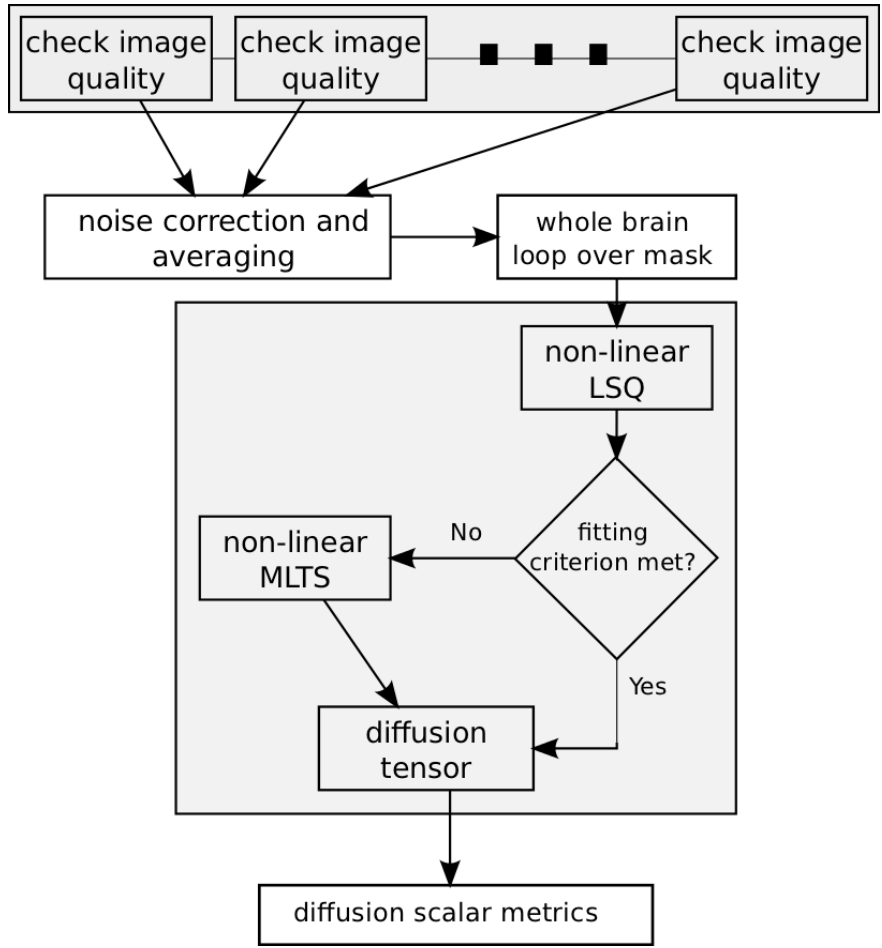

**Figure 9.** Flow diagram of the diffusion imaging framework based on the MLTS algorithm.

The MLTS algorithm consists of following iterative steps (Hawkins and Khan, 2009; Rousseeuw and van Driessen, 2006):

1. Initial guess is obtained from the NLS approach.
2. Residuals are estimated and rearranged using Eq. (5).
3. Truncation factor $h$ is applied.
4. NLS is applied to the truncated function, Eq. (6), using the Levenberg-Marquardt algorithm.
5. Evaluate the fitting criterion: if the criterion is satisfied, stop the iteration, otherwise go to step 2.

Over all estimations we used preselected truncated factor $h$. It should be noted that in the case of a small number of diffusion encoding gradients (less than 16-20) all possible subsets can, in principle, be estimated in order to find the optimal value for $h$ (see the PROGRESS algorithm for details: Rousseeuw and Hubert, 1997). Other approaches (Agullo, 2001; Hofmann et al., 2010) which allow us to determine the optimal truncation factor $h$ are still time consuming and difficult for computation even in the case of parallel estimations. Beside estimation of an optimal truncation factor for a general problem of least trimmed squares computation, we have to take into account that not all subsets can be used for assessment due to *b*-matrix condition constraints. At the moment this problem is outside the scope of this manuscript and demands further investigation. Practically, we can recommend doing a few first trials with *h* between 70% up to 90% of original dataset in order to find the best estimation and fixing it for all further estimations (see some recommendations in Rousseeuw and Van Driessen, 2006; Hawkins and Khan, 2010). In our work we used *h* = 75% both for the simulations and *in vivo* measurements, however, the estimated truncation factor *h* could be increased for some cases with high SNR in order to increase accuracy. Note that significantly under/overestimated truncation factors could lead to biased estimation of diffusion metrics. It means that a selection procedure of truncation factor should be performed carefully.